\documentclass[a4paper,11pt]{article}
\usepackage{pos}
\usepackage{subcaption}

\newcommand{\ket}[1]{\ensuremath{\left|#1\right\rangle}}

\title{CP-violating Dashen phase transition in the two-flavor Schwinger model: a study with matrix product states}
\ShortTitle{CP-violating Dashen phase in the two-flavor Schwinger model}

\author[a,b]{Lena Funcke}
\author[c]{Karl Jansen}
\author*[d]{Stefan Kühn}

\affiliation[a]{Center for Theoretical Physics, Co-Design Center for Quantum Advantage, and NSF AI Institute for Artificial Intelligence and Fundamental Interactions, Massachusetts Institute of Technology, 77 Massachusetts Avenue, Cambridge, MA 02139, USA}
\affiliation[b]{Perimeter Institute for Theoretical Physics, 31 Caroline Street North, Waterloo, ON N2L 2Y5, Canada}
\affiliation[c]{Deutsches Elektronen-Synchrotron DESY, Platanenallee 6, 15738 Zeuthen, Germany}
\affiliation[d]{Computation-Based Science and Technology Research Center,
  The Cyprus Institute, 20 Kavafi Street, 2121 Nicosia, Cyprus}

\emailAdd{lfuncke@mit.edu}
\emailAdd{karl.jansen@desy.de}
\emailAdd{s.kuehn@cyi.ac.cy}

\abstract{We numerically study the Hamiltonian lattice formulation of the two-flavor Schwinger model using matrix product states. Keeping the mass of the first flavor at a fixed positive value, we tune the mass of the second flavor through a range of negative values, thus exploring a regime where conventional Monte Carlo methods suffer from the sign problem and may run into instabilities due to zero modes. Our results indicate a phase transition at the point where the absolute value of the second flavor mass approaches the first flavor mass.  The phase transition is accompanied by the formation of a fermion condensate, a steep drop of the average electric field, and a peak in the bipartite entanglement entropy. Our data hints at a second order transition, which is the 1+1D analog of the CP-violating Dashen phase transition in QCD.\\

Preprint number: MIT-CTP/5348}

\FullConference{%
 The 38th International Symposium on Lattice Field Theory, LATTICE2021
  26th-30th July, 2021
  Zoom/Gather@Massachusetts Institute of Technology
}

\begin{document}
\maketitle

\section{Introduction}

Lattice methods have proven to be invaluable to gain insights into nonperturbative phenomena of gauge theories. In particular, the Markov chain Monte Carlo (MCMC) approach has shed light on many (static) properties, including mass spectra~\cite{Durr2008} and phase diagrams~\cite{Guenther2021}. However, the MCMC approach suffers form the infamous sign problem in certain parameter regimes. A prominent example is the presence of a topological $\theta$-term in QCD, which renders the action complex and thus prevents an efficient Monte Carlo sampling. 

QCD at $\theta=\pi$ is of particular interest, because the theory exhibits a phase transition where CP symmetry gets spontaneously broken. Originally predicted by Dashen based on current algebra arguments~\cite{Dashen1971}, the existence of such a transition can be proven using anomaly matching techniques~\cite{Gaiotto2017}. Due to the anomaly, the regime at $\theta=\pi$ corresponds to having a negative quark mass, which allows for obtaining an intuitive picture for the CP-violating Dashen phase transition. 

\begin{figure}[htp!]
  \centering
  \includegraphics[width=0.4\textwidth]{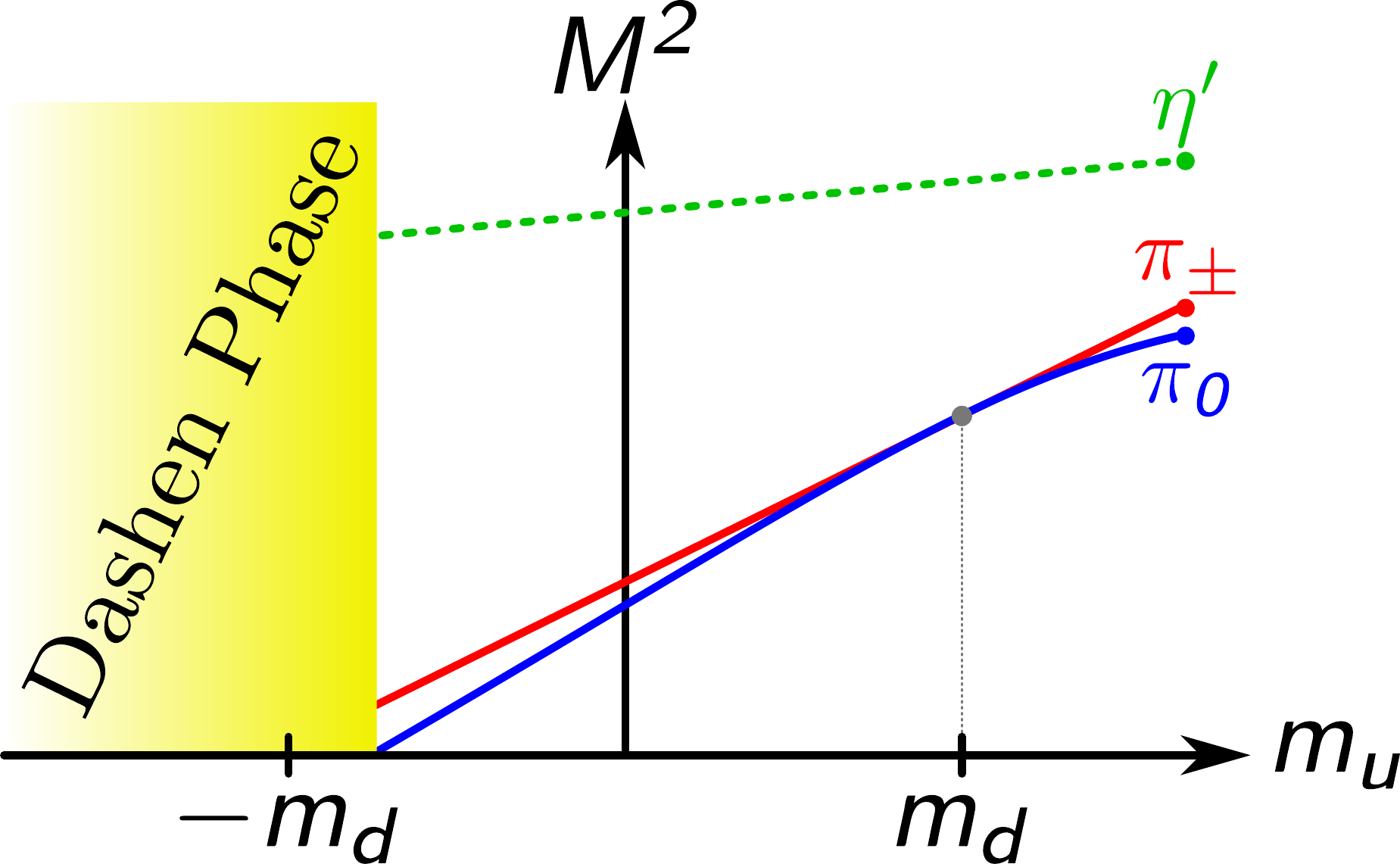}
  \caption{Sketch of the behavior of the low-lying meson spectrum in two-flavor QCD. As the up-quark mass $m_u$ is decreased while keeping the down-quark mass $m_d$ at a fixed positive value, the pion mass $M_{\pi_0}^2\propto m_u+m_d$ eventually becomes complex, signaling the onset of a phase transition. Due to corrections $\propto (m_u-m_d)^2$ this is expected to happen for values of $m_u$ slightly larger than $-m_d$.}
  \label{fig:PhaseDiagram}
\end{figure}

To illustrate this picture, let us focus on QCD with only the lightest two flavors of fermions, the up- and the down-quark, such that the pseudoscalar mesons are given by the pions and the two-flavor analog of the heavy $\eta'$ meson. Neglecting electromagnetic effects, chiral perturbation theory predicts that the square of the pion mass is proportional to the sum of the quark masses, $M_\pi^2\propto m_u + m_d$, where the neutral pion $\pi_0$ has a slightly smaller mass than the charged pions $\pi_\pm$ due to corrections $\propto (m_u-m_d)^2$~\cite{Gasser1983,Gasser1984}. Thus, for a fixed value of $m_d>0$, the theory remains gapped until $m_u\gtrsim -m_d$. Beyond this point, the mass of the neutral pion becomes complex, hence signaling the onset of a phase transition, at which the pion condenses and acquires a vacuum expectation value (see Fig.~\ref{fig:PhaseDiagram}). Since the pion is a CP-odd particle, this transition spontaneously breaks the CP symmetry~\cite{Creutz:1995wf,Creutz2013,Creutz2014}. 

The order of the QCD Dashen phase transition is an open question for the two-flavor case, because it depends on an unknown sign in the effective action~\cite{Creutz:1995wf}. It has been argued that first-order transition lines with second-order end points exist on the $m_d$ axis~\cite{Creutz:1995wf}, which has not yet been studied numerically. We note that the Dashen phase transition plays a crucial role in many models beyond the Standard Model of particle physics (see, e.g., \cite{DiVecchia:2017xpu,Perez:2020dbw}).

In this work, we numerically explore the Dashen phase using the two-flavor Schwinger model as benchmark model, as it shares many similarities with QCD. We study the Hamiltonian lattice formulation of the theory with matrix product states (MPS) and compute the electric field and the analog of the pion condensate in a regime that is inaccessible with conventional MCMC methods. Moreover, we take advantage of the fact that MPS provide a direct access to the entanglement structure in the state and compute the entanglement entropy in the system. This allows us to obtain information about the unknown order of the phase transition.

\section{Model \& Methods}
For our study, we use a Hamiltonian lattice formulation of the Schwinger model with staggered fermions. The dimensionless Hamiltonian for $F$ flavors of fermions on a lattice with $N$ sites and spacing $a$ reads
\begin{align}
  W = &-ix\sum_{n=0}^{N-2}\sum_{f=0}^{F-1}\left(\phi^\dagger_{n,f}e^{i\theta_n}\phi_{n+1,f}-\mathrm{h.c.}\right)+\sum_{n=0}^{N-1}\sum_{f=0}^{F-1}(-1)^n\mu_f\phi^\dagger_{n,f}\phi_{n,f}+ \sum_{n=0}^{N-2} L_n^2.
  \label{eq:Hamiltonian}
\end{align}
In the expression above, the dimensionless parameters $x$ and $\mu_f$ correspond to $x=1/(ag)^2$ and $\mu_f = 2\sqrt{x}m_f/g$, where $g$ is the coupling and $m_f$ the bare fermion mass for flavor $f$. The single-component fermionic fields $\phi_{n,f}$ describe a fermion of flavor $f$ on site $n$, and the operators $L_n$ and $\theta_n$ act on the links between sites $n$ and $n+1$. They are canonically conjugate variables fulfilling $[\theta_m,L_n]=i\delta_{nm}$, and $L_n$ gives the electric flux on the link joining sites $n$ and $n+1$, whereas $e^{i\theta_n}$ acts as a rising operator for the electric flux. We choose to work with a compact formulation, where $\theta_n$ is restricted to $[0,2\pi)$. In addition, the physical states $\ket{\psi}$ of $W$ have to obey Gauss' law, $G_n\ket{\psi} = q_n\ket{\psi}$ $\forall n$, where  
\begin{align}
  G_n = L_n - L_{n-1} - Q_n
  \label{eq:gauss_law}
\end{align}
are the generators of time-independent gauge transformations with the staggered charge
\begin{equation}
Q_n=\sum_{f=0}^{F-1}\phi_{n,f}^\dagger\phi_{n,f}-\frac{F}{2}\left[1-(-1)^n\right],
\end{equation}
and the integer values $q_n$ correspond to static charges. For all the following, we restrict ourselves to the sector of vanishing static charges, $q_n=0$ $\forall n$. For open boundary conditions, Eq.~\eqref{eq:gauss_law} allows us to integrate out the gauge field after fixing the value on the left boundary, which we choose to be zero. When inserting this into the Hamiltonian in Eq.~\eqref{eq:Hamiltonian} and applying a residual gauge transformation, we obtain an expression with only fermionic degrees of freedom given by~\cite{Banuls2013,Banuls2016a,Banuls2016b,Funcke2019}
\begin{align}  
  W' = -ix\sum_{n=0}^{N-2}\sum_{f=0}^{F-1}\left(\phi^\dagger_{n,f}\phi_{n+1,f}-\mathrm{h.c.}\right)
  +\sum_{n=0}^{N-1}\sum_{f=0}^{F-1}(-1)^n \mu_f\phi^\dagger_{n,f}\phi_{n,f} 
  + \sum_{n=0}^{N-2} \left( \sum_{k=0}^nQ_k\right)^2.  
  \label{eq:hamiltonian_dimensionless}
\end{align}
For the rest of the paper, we focus on the case of two fermion flavors. In order to investigate the Dashen phase in the Schwinger model, we fix the bare mass of the first flavor, $m_1/g$, to a positive value and study the vacuum of the theory as we scan the bare mass of the second flavor $m_2/g$ around $-m_1/g$. In the QCD case, the onset of the Dashen phase is characterized by the formation of a pion condensate. Thus, we study the vacuum expectation value of the analog of the pion condensate, which in the continuum theory is given by $\langle\overline{\psi}(x)\gamma^5\tau_3\psi(x)\rangle$, where $\psi(x)$ is a Dirac spinor and $\tau_3$ acts on flavor space and is given by the third Pauli matrix. In units of the coupling, the pion condensate translates to
\begin{align}
  C=i\frac{\sqrt{x}}{N}\sum_{n=0}^{N-2}\sum_{f=0}^{1}(-1)^{n+f}\left(\phi^\dagger_{n,f}\phi^\dagger_{n+1,f}-\mathrm{h.c.}\right)
\end{align}
in our staggered lattice formulation. Moreover, we study the expectation value of the average electric field
\begin{align}
  \bar{F} = \frac{1}{k}\sum_{n=N/2-k/2+1}^{N/2+k/2} L_n,
  \label{eq:average_field}
\end{align}
where we sum over $k$ sites in the center of the system to avoid boundary effects.

In order to compute the ground state of the Hamiltonian numerically, we use the MPS ansatz. For a lattice with $N$ sites and open boundary conditions, the MPS ansatz for the wavefunction reads
\begin{align}
  \ket{\psi} = \sum_{i_0,i_1,\dots,i_{N-1}}A_{i_0}^0 A_{i_1}^1\dots A_{i_{N-1}}^{N-1} \ket{i_0}\otimes \ket{i_1}\otimes \dots\otimes \ket{i_{N-1}},
  \label{eq:mps}
\end{align}
where the $A_{i_n}^n$ are complex $D\times D$ matrices for $0<n<N-1$, and $A_{i_0}^0$ ($A_{i_{N-1}}^{N-1}$) is a $D$-dimensional row (column) vector. The set of states $\{\ket{i_{n}}\}_{n=0}^{d-1}$ forms a basis for the $d$-dimensional Hilbert space at site $n$. The parameter $D$, called the bond dimension of the MPS, determines the number of the variational parameters in the ansatz and limits the maximum amount of entanglement that can be present in the state (see Ref.~\cite{Schollwoeck2011} for a detailed review). In particular, MPS and tensor networks in general allow for reliable computations in regimes where conventional MCMC methods suffer from the sign problem~\cite{Banuls2016a,Banuls2018a,Banuls2019,Felser2019,Magnifico2021,Nakayama2021}. Moreover, one has direct access to the entanglement structure in the state, which allows us to study the von Neumann entropy $S$ for the reduced density matrix that describes the first $N/2$ sites of the system. Although tensor networks can directly deal with fermionic degrees of freedom, we choose to translate them to spins using a Jordan-Wiger transformation for convenience in the numerical simulations~\cite{Banuls2016a,Banuls2018a,Funcke2019}.

\section{Results}
\begin{figure}[htp!]
  \centering
  \includegraphics[width=0.98\textwidth]{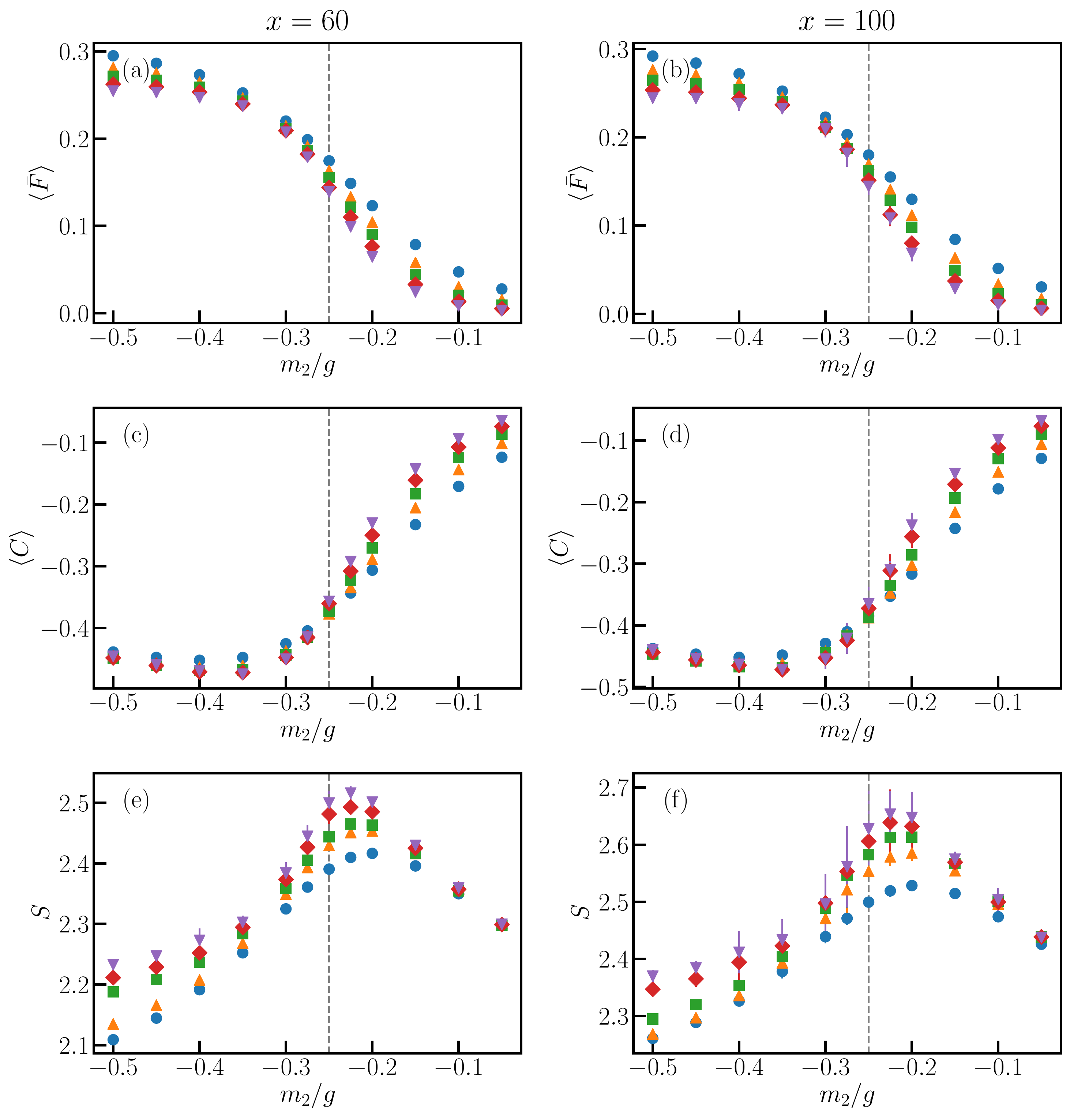}
  \caption{Average electric field (first row), pion condensate (second row) and entropy (third row) as a function of $m_2/g$ for $m_1/g=0.25$, $x=60$ (first column) and $x=100$ (second column). The different markers indicate data for different volumes $N/\sqrt{x}=10$ (blue dots), $12.5$ (orange triangles), $15$ (green squares), $17.5$ (red diamonds) and $20$ (purple upside-down triangle). The error bars arise from the extrapolation on $D$. The dashed vertical line indicates the point where $m_2/g$ reaches $-m_1/g$. To compute the average electric field we use $k=4$ sites in the center of the system according to Eq.~\eqref{eq:average_field}.}
  \label{fig:res_extD}
\end{figure}

In our study, we set $m_1/g$ to $0.25$ and use fixed dimensionless physical volumes $N/\sqrt{x}$ ranging from $10$ to $20$, where the lattice spacing corresponds to $x\in[60;100]$. Moreover, we choose to work in the sector of vanishing total charge, $\sum_n Q_n=0$. Compared to a conventional lattice calculation, we have an additional source of error due to the limited matrix size $D$ that can be reached in our numerical simulations. In order to control this error, we repeat our simulation for every combination of $(N/\sqrt{x}, x, m_1/g, m_2/g)$ for several values of $D$ and extrapolate our results to the limit $D\to\infty$ following the procedure in Ref.~\cite{Funcke2019}. Our results for the average electric field, the pion condensate, and the entropy in the vacuum as a function of $m_2/g$ are shown in Fig.~\ref{fig:res_extD}.

For the average electric field in Figs.~\ref{fig:res_extD}(a) and \ref{fig:res_extD}(b), we observe a drop from nonvanishing values at $m_2/g=-0.5$ to values close to zero as $m_2/g$ approaches $-m_1/g$. A comparison between our results for the coarsest and the finest lattice spacing (corresponding to $x=60$ and $x=100$, respectively, see Figs.~\ref{fig:res_extD}(a) and \ref{fig:res_extD}(b)) shows that there is no strong $x$ dependence throughout the entire parameter regime we study. In contrast, there is a mild dependence on the physical volume: with increasing volume the drop in the average electric field becomes slightly sharper as $m_2/g$ approaches larger values, and the final value of the field is closer to zero as $m_2/g$ goes towards zero. 

The regime with the negative fermion mass corresponds to the presence of a topological term with angle $\theta=\pi$~\cite{Creutz2013}. The single-flavor Schwinger model is known to undergo a phase transition at some critical mass in this regime. It has been observed that the electric field vanishes before the transition and has a nonvanishing value after the transition point~\cite{Byrnes2002,Buyens2017}. Our results for the two-flavor case are compatible with this observation, and the drop in the average electric field hints towards a phase transition at $m_2/g\approx -m_1/g$.

For the pion condensate (see Figs.~\ref{fig:res_extD}(c) and \ref{fig:res_extD}(d)), we see a similar picture as for the average electric field. For large values of $m_2/g$, the value of the condensate is close to zero. It decreases as we get closer to $-m_1/g$ and eventually approaches a constant value upon further lowering the value of $m_2/g$. Again, the dependence on the lattice spacing is negligible for the parameter range we study. The effect of the finite volume is slightly larger for the pion condensate than for the electric field. In the regime of large values of $m_2/g$ close to zero, we observe that the condensate value does not vanish completely but gets closer to zero as we increase the volume. 
These observations for the pion condensate are compatible with the expectation for the Dashen phase in QCD with two flavors of fermions, see Fig.~\ref{fig:PhaseDiagram}. For $m_2/g\gg -m_1/g$, we are outside of the Dashen phase, and the values for the pion condensate are close to zero. As we approach $m_2/g\approx -m_1/g$, we enter the Dashen phase, and the values of the condensate decrease. Eventually, we obtain approximately constant nonvanishing values for $m_2/g\ll -m_1/g$. Thus, the behavior of the condensate confirms the occurrence of the Dashen phase transition at $m_2/g\approx -m_1/g$.

Finally, we can also look at the entanglement entropy shown in Figs.~\ref{fig:res_extD}(e) and \ref{fig:res_extD}(f) for $x=60$ and $x=100$. Compared to the average electric field and the pion condensate, the entropy shows a more pronounced dependence on the volume and the lattice spacing, especially for small values of $m_2/g$, and a clear peak around $m_2/g=-0.225$. In particular, the volume dependence of the peak is expected for a second (or higher order) phase transition. The entropy is directly related to the correlation length in the system, which diverges logarithmically (in the thermodynamic limit) as one approaches the phase transition~\cite{Vidal2003a,Latorre2004,Calabrese2009}. Since we are working with a finite system, and the system size upper bounds the correlation length, we expect $S$ to diverge logarithmically with the physical volume for a fixed value of $x$, $S\propto\log(N/\sqrt{x})$. In contrast, the correlation length is finite if we go away from the critical point, and the entropy should eventually saturate upon reaching large enough volumes.
\begin{figure}[htp!]
  \centering
  \includegraphics[width=0.98\textwidth]{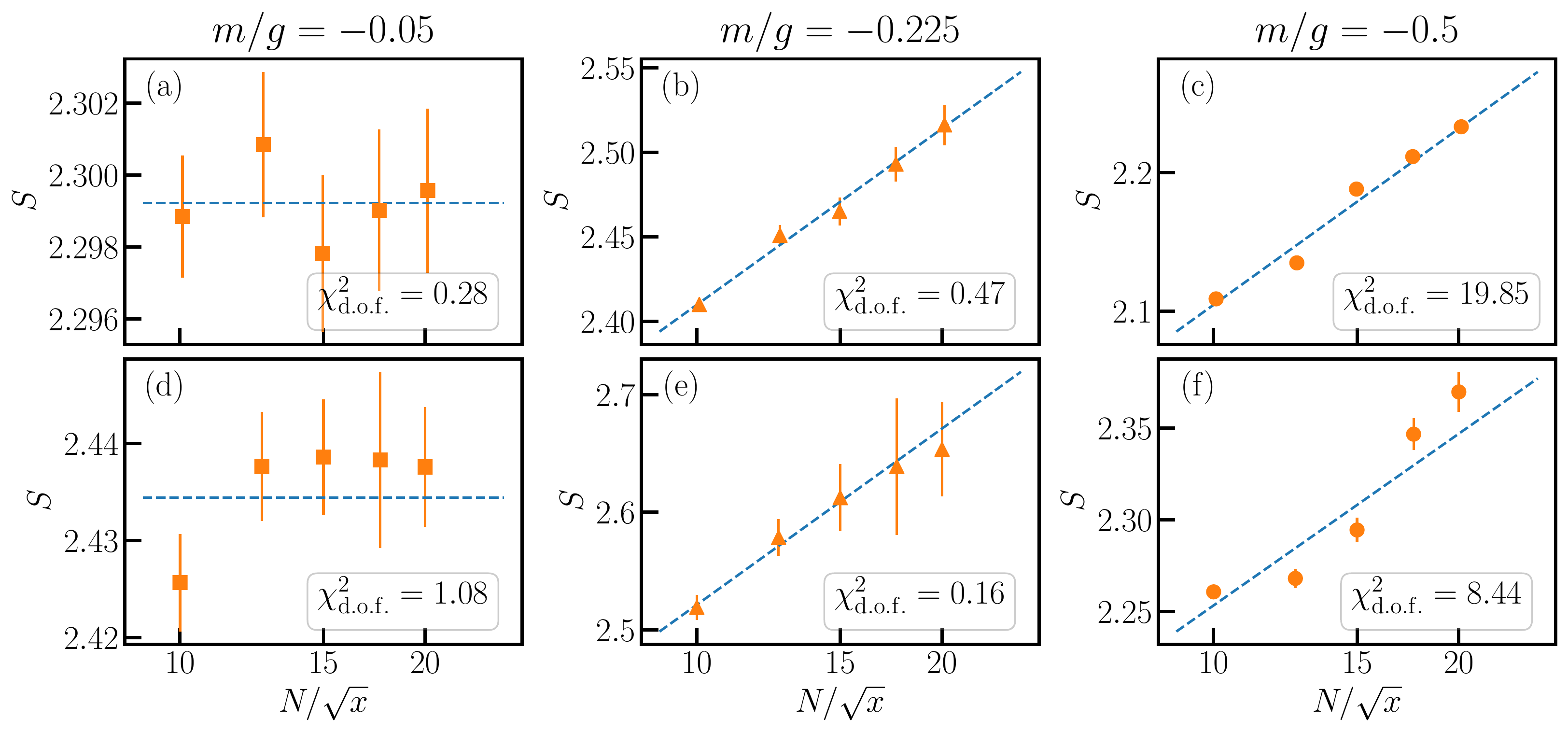}
  \caption{Entropy as a function of volume for $x=60$ (top row) and $x=100$ (bottom row). The different columns correspond to $m_2/g=-0.05$ (first column), $m_2/g=-0.225$ (middle column), and $m_2/g=-0.5$ (right column). Note that the $x$-axis is in logarithmic scale. The blue dashed lines in panels (a) and (d) correspond to a constant fit, whereas in on all other panels it corresponds to a fit $S(N/\sqrt{x}) = c_1\log(N/\sqrt{x}) + c_2$ with constants $c_1$, $c_2$, and the insets show the reduced $\chi^2$ of the fit.}
  \label{fig:entropy_vs_volume}
\end{figure}
The dependence of the entropy as a function of volume for various values of $m_2/g$ is shown in Fig.~\ref{fig:entropy_vs_volume}. For our largest value of $m_2/g$, $-0.05$, we indeed observe that the entropy is essentially constant as a function of $N/\sqrt{x}$ (see Figs.~\ref{fig:entropy_vs_volume}(a) and \ref{fig:entropy_vs_volume}(d)). 
For the opposite limit of $m_2/g=-0.5$, our data for the entropy are to a certain extent compatible with a logarithmic divergence at first glance.
Focusing on $x=60$ first (see Fig.~\ref{fig:entropy_vs_volume}(c)), for which our numerical data is most precise, fitting our data in that regime to a logarithmic divergence yields relatively large values of $\chi^2_\mathrm{d.o.f.}$. In particular, we also observe a noticeable change in the scaling behavior as we exceed a volume of $12.5$. Together with the value of $\chi^2_\mathrm{d.o.f.}$ this indicates that our results for $m_2/g=-0.5$ are not very well described by a logarithmic divergence. 
Our data for $x=100$ (cf. Fig.~\ref{fig:entropy_vs_volume}(f)) show a similar behavior albeit being less precise due to the larger values of $N$ required to reach the same physical volumes. In contrast, for $m_2/g=-0.225$ our results are very well compatible with a logarithmic divergence in volume, as Figs.~\ref{fig:entropy_vs_volume}(b) and \ref{fig:entropy_vs_volume}(e) reveal.

\section{Discussion \& Outlook}
In summary, we used MPS to explore the Dashen phase of the two-flavor Schwinger model, which is a regime that is inaccessible with conventional MCMC methods. Fixing the bare mass $m_1/g$ of the first flavor to a positive value and scanning the bare mass $m_2/g$ of the second flavor through a range of negative values, we observe clear indications for a phase transition around the point where the absolute values of both masses are equal. We observe steep drops of the average electric field and the pion condensate. 
In particular, our data for the pion condensate suggest that the observed transition is indeed the 1+1D analog of the CP-violating Dashen phase transition in QCD. 

When studying the bipartite entanglement entropy, we observe a clear peak at a value of $m_2/g\approx -0.225$, which is slightly larger than $-m_1/g=-0.25$. This indicates that the onset of the Dashen phase happens not exactly at $-m_1/g=m_2/g$ but at $-m_1/g<m_2/g$, in agreement with the theoretical expectation. In two-flavor QCD, the pion mass gets corrections $\propto (m_u-m_d)^2$, which shift the Dashen phase transition to larger masses of the down-quark~\cite{Gasser1983,Gasser1984,Creutz2013,Creutz2014,Creutz2019}. 

In addition, we also examined the scaling behavior of the entropy with the volume for a fixed lattice spacing. At the transition point, our data are compatible with a logarithmic divergence in the volume, which suggests that the observed transition is of second (or higher) order. This agrees with the theoretical expectation \cite{Creutz:1995wf} that two first-order transition lines with second-order end points should exist on the $m_d$ axis, which had not been numerically verified so far. 
In the future, we aim to unambiguously pinpoint the order of the transition by increasing the accuracy, both  with larger values of the bond dimension and with larger physical volumes. 

\acknowledgments
S.K.\ acknowledges financial support from the Cyprus Research and Innovation Foundation under project ``Future-proofing Scientific Applications for the Supercomputers of Tomorrow (FAST)'', contract no.\ COMPLEMENTARY/0916/0048. 
L.F.\ is partially supported by the U.S.\ Department of Energy, Office of Science, National Quantum Information Science Research Centers, Co-design Center for Quantum Advantage (C$^2$QA) under contract number DE-SC0012704, by the DOE QuantiSED Consortium under subcontract number 675352, by the National Science Foundation under Cooperative Agreement PHY-2019786 (The NSF AI Institute for Artificial Intelligence and Fundamental Interactions, \url{http://iaifi.org/}), and by the U.S.\ Department of Energy, Office of Science, Office of Nuclear Physics under grant contract numbers DE-SC0011090 and DE-SC0021006.
Research at Perimeter Institute is supported in part by the Government of Canada through the Department of Innovation, Science and Industry Canada and by the Province of Ontario through the Ministry of Colleges and Universities. 

\bibliographystyle{JHEP}
\bibliography{Papers}

\end{document}